# Measurement of the Integrated Faraday Rotations of BL Lac Objects

A. B. Pushkarev

*Astro-Space Center, Lebedev Physical Institute, Leninskiĭ pr. 53, 117924 Russia*

**Abstract**—We present the results of multi-frequency polarization VLA observations of radio sources from the complete sample of northern, radio-bright BL Lac objects compiled by H. Kühr and G. Schmidt. These were used to determine the integrated rotation measures of 18 sources, 15 of which had never been measured previously, which hindered analysis of the intrinsic polarization properties of objects in the complete sample. These measurements make it possible to correct the observed orientations of the linear polarizations of these sources for the effect of Faraday rotation. The most probable origin for Faraday rotation in these objects is the Galactic interstellar medium. The results presented complete measurements of the integrated rotation measures for all 34 sources in the complete sample of BL Lac objects.

## 1. INTRODUCTION

BL Lac objects are an extreme class of active galactic nuclei characterized by a number of peculiar properties: weak or absent optical line emission, rapid and strong variability in all parts of the electromagnetic spectrum, a high degree of linear polarization, and compact radio and pointlike optical structure.

In 1990, Kühr and Schmidt [1] compiled a complete sample of radio-bright BL Lac objects in the Northern sky, consisting of 34 sources satisfying the following selection criteria: (1) total flux at 5 GHz $I \geq 1$ Jy at the epoch at which the sample was compiled, (2) two-point spectral index between 11 and 6 cm $\alpha_{11-6} \geq -0.5$ ($S \sim \nu^{+\alpha}$), and (3) equivalent width of emission lines $W \leq 5$ Å. The sample included sources brighter than $20^m$ in the optical with declinations $\delta > 20°$ located outside the Galactic plane ($|b| > 10°$). We are currently analyzing multi-frequency, polarization-sensitive Very Long Baseline Interferometry (VLBI) data for all sources in the sample.

VLBI polarization observations can yield radio images of compact sources such as BL Lac objects with angular resolution of the order of a millisecond of arc. Such observations contain information about all four Stokes parameters, i.e., both the total intensity and the linear and circular polarization of the radiation. The emission of BL Lac objects is characteristically linearly polarized, since it is synchrotron radiation. Thus, polarization-sensitive observations carry information not only about the energetics and kinematics of the sources observed, but also about the degree of order and orientation of the magnetic fields in their jets, which are very important when analyzing the physical conditions in these objects.

There are a number of factors that can complicate deduction of the direction of the magnetic field in a particular region in a source. Three properties of the region must be known: first, the orientation of the electric vector of its linear polarization, second its optical depth, and finally, its rotation measure.

The first two properties flow directly from any multi-frequency polarization observations, and yield the orientation of the magnetic field in the radiating region relative to the observed polarization electric vector. When a sufficiently high degree of polarization is detected at even one frequency, this indicates that the radiating plasma must be optically thin, since optically thick radiation cannot be more than $\simeq 10$–15% polarized [2]. If the plasma is optically thin, the magnetic and electric fields are mutually perpendicular, whereas the two fields are parallel if the plasma is optically thick. The core of the source is usually considered to be optically thick, while the jet becomes increasingly transparent—optically thin—with distance from the core. This is demonstrated by the presence of highly polarized jet components [3] (isolated regions of brightening of the total intensity), whose degree of polarization can reach 60–70%—the limit for optically thin synchrotron radiation [2]. The high degrees of polarization that are sometimes observed in the VLBI jets of BL Lac objects testify that the magnetic fields are well ordered in these regions. As a rule, the jet emission has a steep spectrum, confirming that we are dealing here with an optically thin plasma.

Faraday rotation of the plane of linear polarization occurs when polarized radiation passes through a region containing thermal plasma and a magnetic field with a non-zero component along the line of sight. The rotation measure can only be derived from multi-frequency polarization observations. However, wide-band polarimetry (1.4, 5, 8, 15, 22 GHz, for example) is not suited to this purpose. As is discussed in [5], the best suited observations are those made with the NRAO VLA at 1.3–1.7 GHz. When planning observations designed to derive the rotation measures of compact extragalactic sources, several requirements on the frequency interval to be used should be taken into consideration. First, there should be a large enough spacing between the corresponding wavelengths (more precisely, the squares of the wavelengths) to detect the rotation of the plane of linear polarization. Second, these wavelengths should be long enough not to give rise to large ambiguities when calculating the rotation measures. Finally, the wavelengths should not be very different from each other, since otherwise, there is the possibility of detecting radiation from electrons with qualitatively different energies—i.e., from physical different regions in the source—at the different wavelengths.

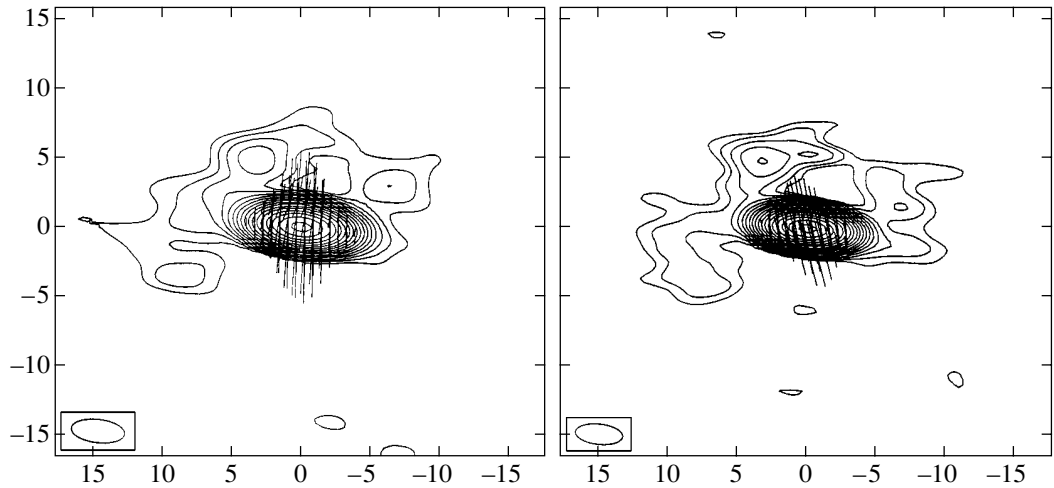

**Fig. 1.** Total-intensity maps of 0119 + 115 at 1.41 GHz (left) and 1.63 GHz (right) with linear polarization vectors superposed. The peak flux densities are 892 and 891 mJy, respectively, and the lowest contours are 0.35% left and 0.25% right of the peaks, respectively. The horizontal and vertical scales are in arcseconds.

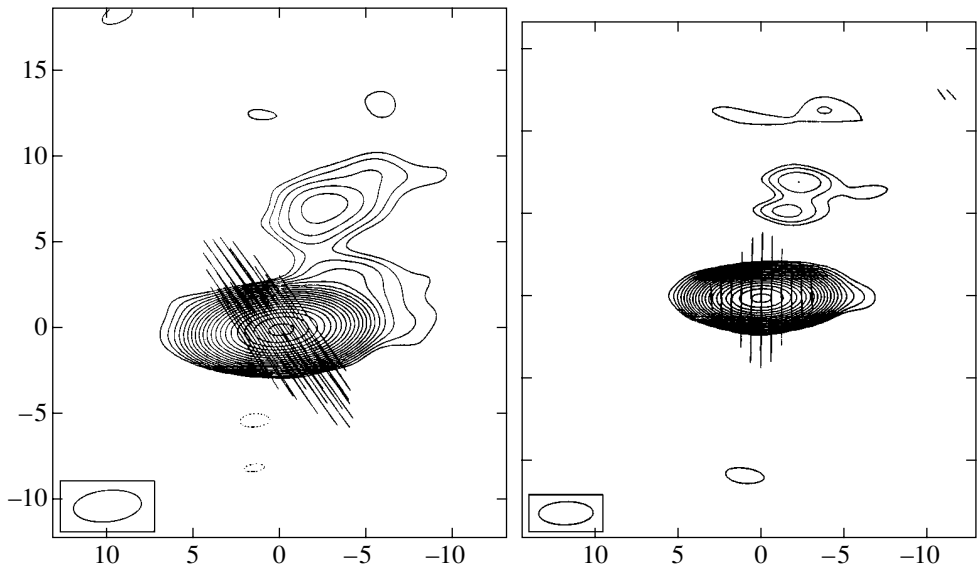

**Fig. 2.** Total-intensity maps of 0235 + 164 at 1.41 GHz (left) and 1.66 GHz (right) with linear polarization vectors superposed. The peak flux densities are 833 and 1032 mJy, respectively, and the lowest contours are 0.125% of the peaks. The horizontal and vertical scales are in arcseconds.

## 2. OBSERVATIONS

We obtained VLA observations (C or D configuration) during four sessions: January 19, 1996 (Session 1), January 29, 1996 (Session 2), February 7, 1996 (Session 3), and June 26, 1998 (Session 4). The 1998 observations were conducted in order to refine our initial determination of the rotation measure for 0119 + 115, since the 1996 data for this source were of comparatively low quality, leading to a large uncertainty in its rotation measure. All the observations were carried out at six frequencies (1664.9, 1635.1, 1515.9, 1467.7, 1417.5, and 1365.1 MHz), with bandwidths of 50 MHz for the observations at 1664.9 and 1635.1 MHz and 25 MHz for the remaining frequencies. The data were calibrated in the AIPS (Astronomical Image Processing System, NRAO) package. The primary flux calibrators were 3C 286 (Sessions 2 and 3) and 3C 48 (Sessions 1 and 4). The polarization position angles were calibrated using 3C 286 (Sessions 2 and 3) and 3C 138 (Sessions 1 and 4), for which we assumed the polarization position angles $\chi = +33°$ and $\chi = -9°$, respectively, at all frequencies, since these sources do not have significant rotation measures [8]. The instrumental polarizations were determined using sources with observations spanning a large interval (>90°) in parallactic angle—2150 + 173 (Session 1), 0917 + 624 (Session 2), 1147 + 245 (Session 3), and 0235 + 495 (Session 4). The unpolarized sources 2352 + 495 (Session 1, $m_L \leq 1\%$), 1323 + 321 (Sessions 2 and 3, $m_L \leq 0.6\%$), and 0134 + 329 (Session 4, $m_L \leq 0.8\%$) were used to check the results of the solutions for the instrumental polarizations.

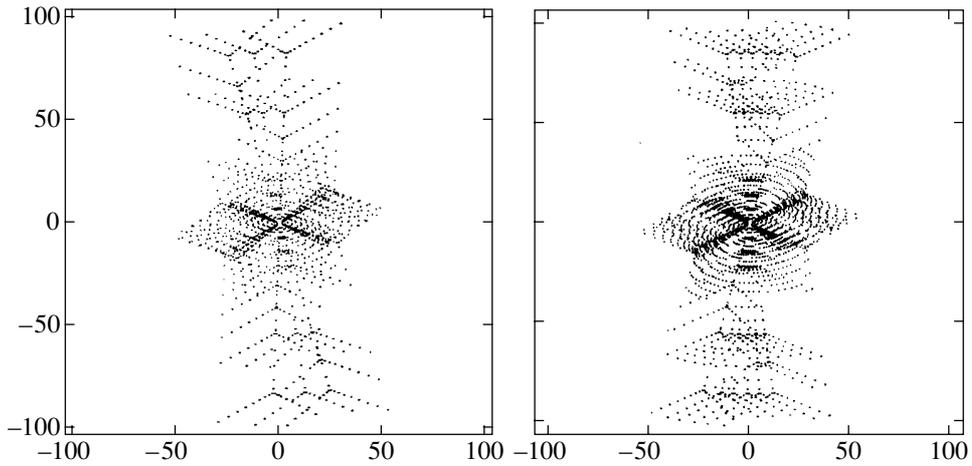

**Fig. 3.** Coverage of the *u–v* plane for the VLA observations of 0119 + 115 (left, three scans) and 0235 + 164 (right, six scans). The horizontal and vertical scales are thousands of wavelengths.

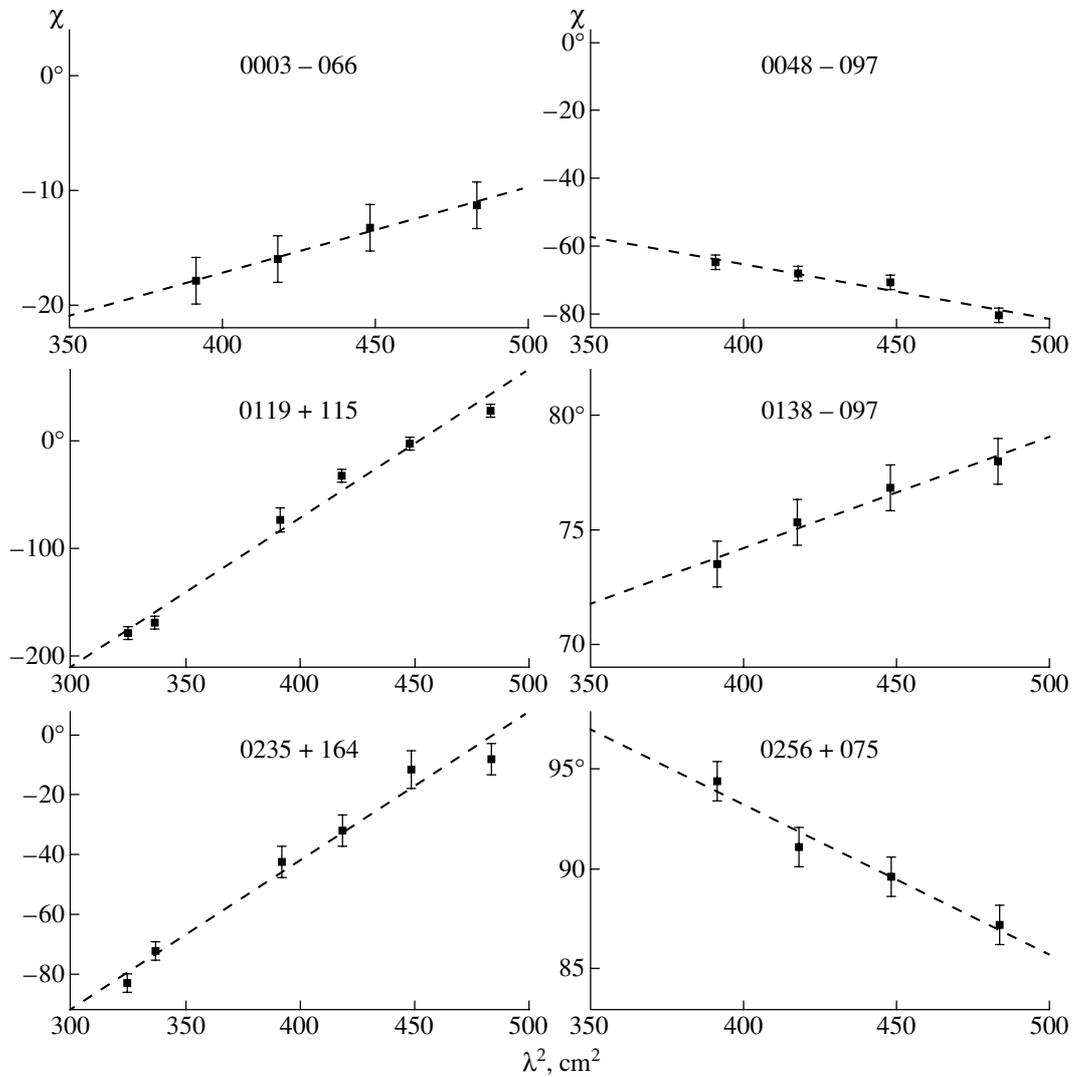

**Fig. 4.** Dependence of the polarization position angle $\chi$ (in degrees) on the square of the wavelength (in cm$^2$) for the sample sources.

## 3. RESULTS

Structure on VLA (kiloparsec) scales was detected in the two sources 0119 + 115 and 0235 + 164. Figures 1 and 2 show images of the distributions of total intensity for these sources, with linear-polarization vectors superposed. The remaining sources were unresolved at the sensitivity level reached in our observations, and so their images are not shown. The images in Figs. 1 and 2 were made at different frequencies, and variations in the polarization angles are clearly visible. The orientation of these vectors corresponds to that of the polarization electric vectors, and their length is proportional to the degree of polarization. The sizes and shapes of the antenna beams are indicated in the lower left corner of each image. Figure 3 shows the $u$–$v$ coverages for 0119 + 115 and 0235 + 164 for our observations, which are typical for VLA "snap-shot" observations of sources at these declinations.

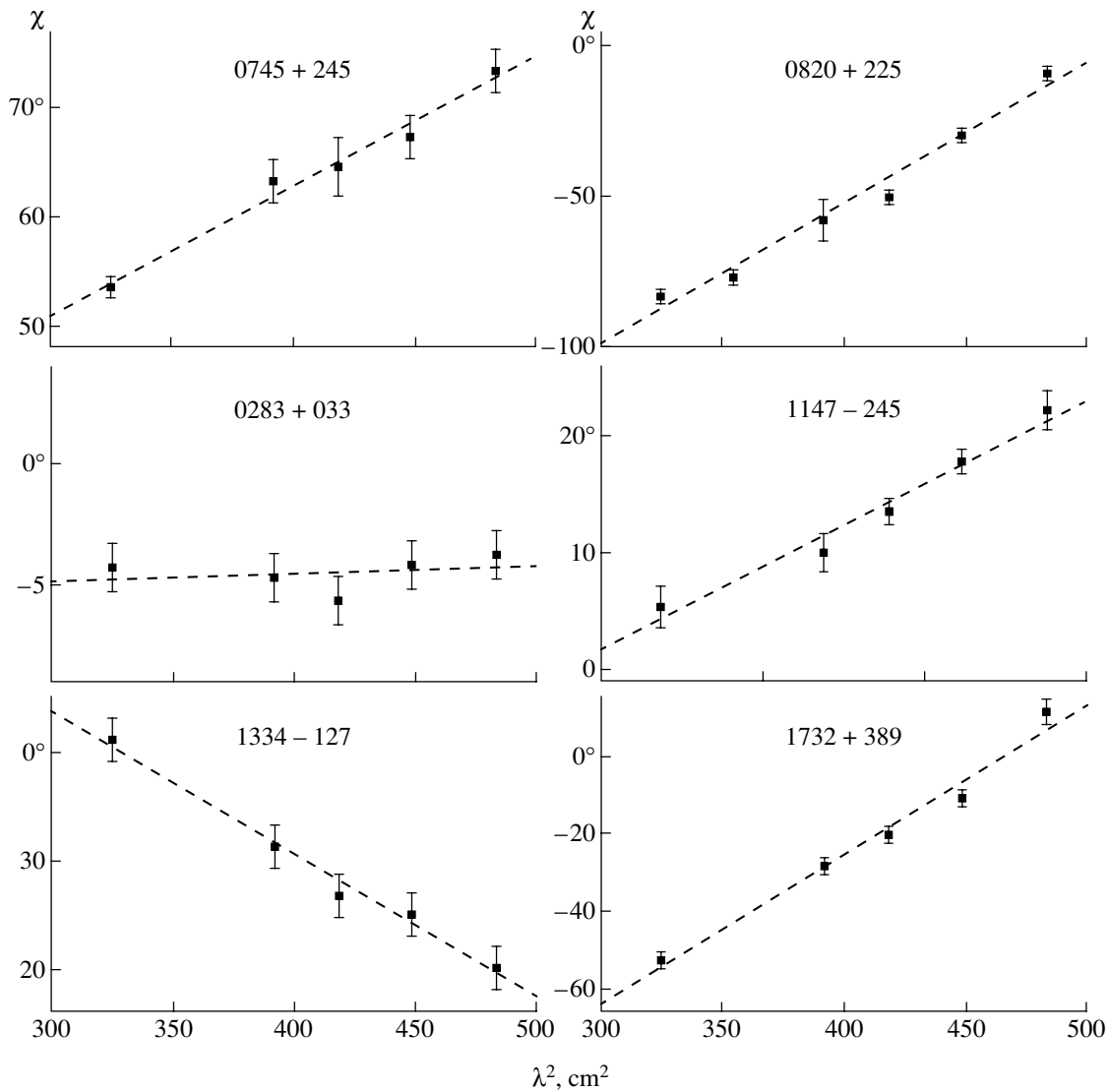

**Fig. 4.** (Contd.)

Figure 4 presents the dependence of the polarization position angle $\chi$ on the square of the wavelength for the 18 sources. In all cases, these plots show a linear dependence on $\lambda^2$. The desired rotation measure is the slope of the line

$$\chi(\lambda^2) = \chi_0 + \text{RM}\lambda^2, \quad \text{RM} = 0.81 \int N_e B \cos\theta \, dl, \quad (1)$$

where $N_e$ is the density of thermal electrons (cm$^{-3}$), $B$ the magnetic field (µG), $\theta$ the angle between the direction of propagation of the radiation and the magnetic-field direction, and $dl$ an element of path length from the source to the observer (pc). This expression for the rotation measure is valid for quasi-longitudinal propagation (so that the angle $\theta$ is small) and a weak magnetic field [4].

As a rule, errors in $\chi$ are several degrees, and are presented at the 2$\sigma$ level; we had several observations for each source at each frequency, enabling us to estimate

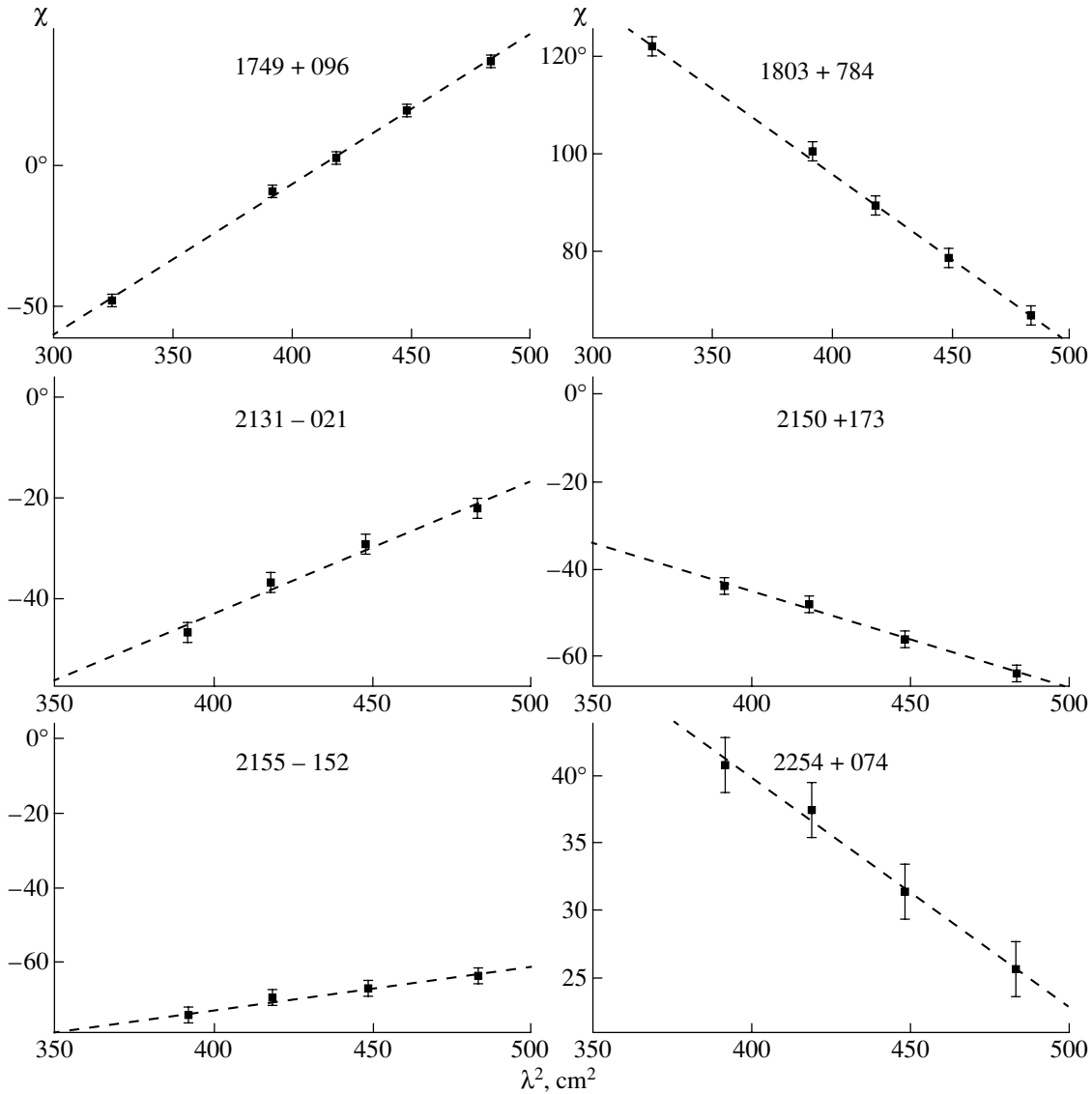

**Fig. 4.** (Contd.)

Integrated rotation measures for all sources in the complete sample of BL Lac objects the errors in $\chi$ by comparing these independent measurements. We determined linear least-squares fits to the observed $\chi$ dependences. Note that the polarization angle is determined with accuracy to within $\pm n\pi$ ($n = 0, 1, 2, \ldots$), which can lead to ambiguity in the rotation measure RM, since we are dealing with observations at a discrete set of frequencies. The square of the minimum difference between neighboring wavelengths will determine this ambiguity, which in our case is $\pm 1190$ rad/m$^2$. Nevertheless, the presence of six observing frequencies and their proximity to one another enables us to determine the rotation measures with certainty.

| Source | Rudnick and Jones [5] | Rusk [6] | Wrobel [7] | Our observations | $z$ | Scale, Kpc/arcsec[1] |
|---|---|---|---|---|---|---|
| 0003 − 066 | | | | 13 ± 2 | 0.347 | 2.9 |
| 0048 − 097 | | | | −28 ± 7 | – | – |
| 0119 + 115 | | | | 240 ± 18 | 0.570 | 3.7 |
| 0138 − 097 | | | | 9 ± 2 | 0.440 | 3.3 |
| 0235 + 164 | 57 ± 20 | 57 ± 21 | | 85 ± 9 | 0.940 | 4.2 |
| 0256 + 075 | | | | −13 ± 2 | 0.893 | 4.2 |
| 0454 + 844 | | 5 ± 6 | | | 0.112 | 1.3 |
| 0716 + 714 | | −30 ± 3 | | | – | – |
| 0735 + 178 | 9 ± 13 | 9 ± 12 | | | 0.424 | 3.3 |
| 0745 + 241 | | | | 21 ± 2 | 0.410 | 3.2 |
| 0814 + 425 | | 19 ± 9 | | | 0.245 | 2.4 |
| 0820 + 225 | | | | 81 ± 10 | 0.951 | 4.2 |
| 0823 + 033 | | | | 1 ± 2 | 0.506 | 3.6 |
| 0828 + 493 | | −6 ± 5 | | | 0.548 | 3.7 |
| 0851 + 202 | 31 ± 2 | 31 ± 3 | | | 0.306 | 2.8 |
| 0954 + 658 | | −15 ± 4 | | | 0.368 | 3.1 |
| 1147 + 245 | | | | 19 ± 2 | – | – |
| 1308 + 326 | −4 ± 4 | −4 ± 4 | | | 0.996 | 4.3 |
| 1334 − 127 | | | | −23 ± 3 | 0.539 | 3.7 |
| 1418 + 546 | 17 ± 7 | | | | 0.151 | 1.7 |
| 1538 + 149 | | 17 ± 2 | | | 0.605 | 3.8 |
| 1652 + 398 | | 42[2] | | | 0.033 | 0.5 |
| 1732 + 389 | | | | 67 ± 8 | 0.970 | 4.2 |
| 1749 + 096 | 105 ± 22 | | | 94 ± 6 | 0.322 | 3.0 |
| 1749 + 701 | 33 ± 22 | 16 ± 5 | 8 ± 13 | | 0.770 | 4.0 |
| 1803 + 784 | | −70 ± 5 | −61 ± 4 | −61 ± 2 | 0.680 | 3.9 |
| 1807 + 698 | 20 ± 46 | 5 ± 5 | 14 ± 4 | | 0.051 | 0.7 |
| 1823 + 568 | | 34 ± 5 | 36 ± 4 | | 0.664 | 3.9 |
| 2007 + 777 | | −20 ± 3 | | | 0.342 | 3.1 |
| 2131 − 021 | | | | 46 ± 7 | 1.285 | 4.3 |
| 2150 + 173 | | | | −39 ± 4 | – | – |
| 2155 − 152 | | | | 19 ± 4 | 0.672 | 3.9 |
| 2200 + 420 | −205 ± 6 | | | | 0.068 | 0.8 |
| 2254 + 074 | | | | −30 ± 3 | 0.190 | 2.0 |

[1] For a uniformly expanding Universe ($q_0 = 0.5$) with $H_0 = 100$ km s$^{-1}$ Mpc$^{-1}$.
[2] Author does not present errors for this measurement.

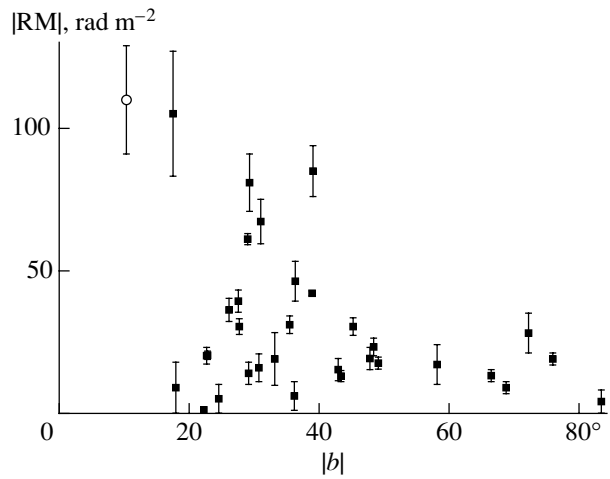

**Fig. 5.** Dependence of the absolute value of the rotation measure on the absolute value of the Galactic latitude. For sources with several RM measurements, we chose the RM value obtained at the latest epoch. The new value for the Galactic contribution to the rotation measure of BL Lac found by Reynolds *et al.* [9] is shown by the hollow circle.

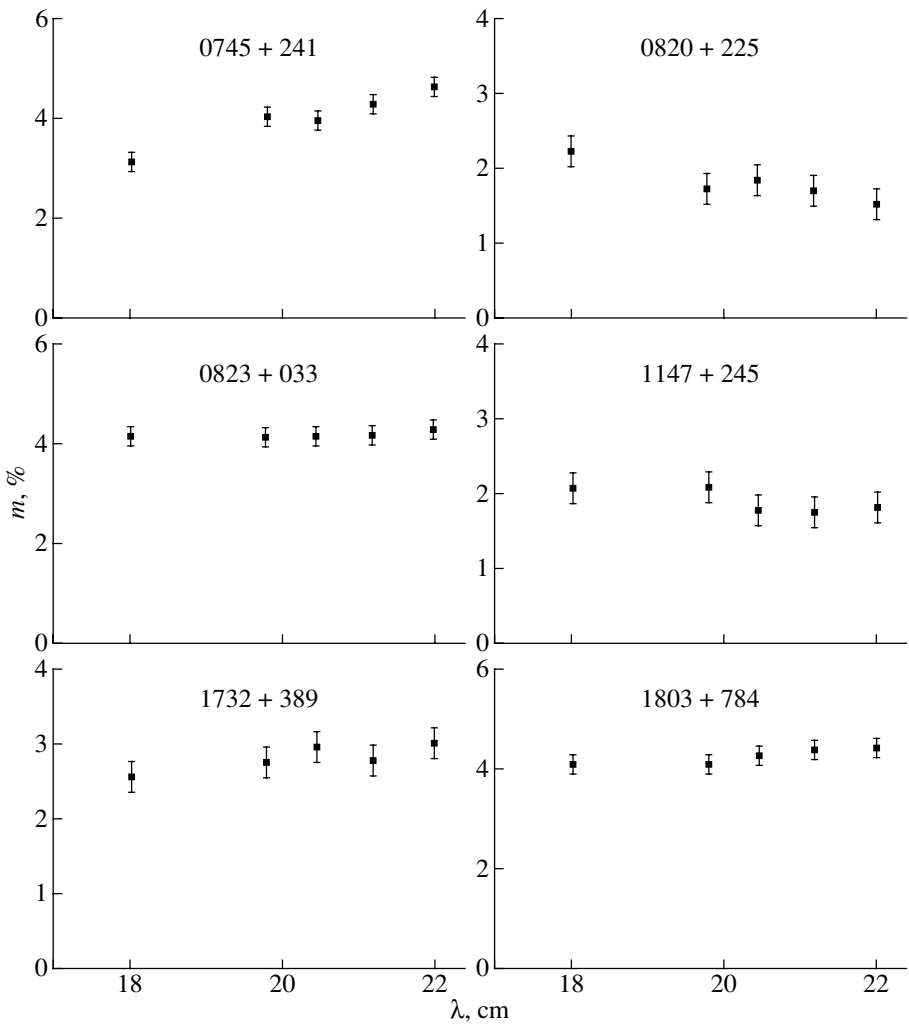

**Fig. 6.** Dependence of the degree of polarization on wavelength using six sources as examples. There is no evidence for depolarization.

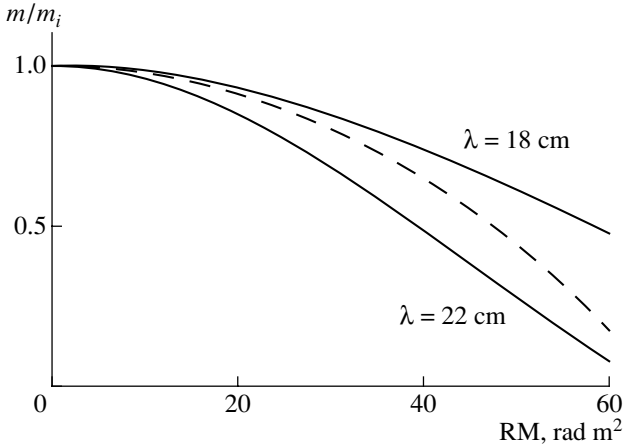

**Fig. 7.** Dependence of the relative degree of polarization $m/m_i$ at various wavelengths on the rotation measure, where $m_i$ is the intrinsic degree of polarization in the absence of internal Faraday rotation and $m$ is the degree of polarization expected for a plane-layer model [9] in which a substantial fraction of the rotation measure is internal to the source. The dashed curve shows the theoretical dependence of the ratio of the degrees of polarization $m_{\lambda_1}/m_{\lambda_2}$ for wavelengths $\lambda\lambda$ 18 and 22 cm predicted by this model.

The integrated rotation measures for the 18 sources from the complete sample of Kühr and Schmidt [1] are presented in the table, together with results for the remaining 16 sample sources obtained in earlier observations [5–7]. These values are primarily in the interval $|RM| = 0$–$100$ rad/m$^2$, which coincide with the Galactic rotation measures derived from observations of 555 radio sources [8]. Only two objects have $|RM| > 200$ rad/m$^2$. It is probable that the Faraday rotation we have detected occurs primarily when the radiation passes through the interstellar medium of the Galaxy. Further evidence for this is provided by the distribution of the $|RM|$ values for all the sample sources relative to the absolute value of the Galactic latitude, where we can see a tendency for $|RM|$ to decrease with increase $|b|$ (Fig. 5).

Figure 5 does not show the two highest values of $|RM|$, for the sources 0119 + 115 and 2200 + 420 (BL Lac). Results obtained by Reynolds *et al.* [9] (personal communication) based on multi-frequency VLBI observations show that the Galactic contribution to the observed integrated rotation measure of BL Lac is $-110 \pm 19$ rad/m$^2$, indicated in Fig. 5 by the hollow circle. We can see that this new value agrees well with the general tendency shown by the remaining rotation measures of the sample sources. It is not ruled out that the rotation measure for 0119 + 115 also includes an appreciable contribution from regions near the source.

In addition, we do not detect evidence for depolarization (Fig. 6); the degree of polarization remains essentially constant with frequency. Depolarization arises when the thermal plasma associated with the RM is mixed in with the source [10]. In this case, the polarizations from the far and near regions of the source add incoherently, and the resulting observed polarized decreases as the Faraday depth increases. This is shown for a plane-layer model [10] in Fig. 7, which presents theoretical depolarization curves for various wavelengths when an appreciable fraction of the rotation measure is internal to the source.

Moreover, another piece of evidence supporting a Galactic origin for most of the observed rotation measures is that the RM values apparently do not vary with time, as indicated by the values in the table for sources for which there are observations at more than one epoch; note that all the results in the table were obtained using the same observing and reduction techniques. In addition, as was shown in [11], the rotation measures remain constant over periods of decades, in spite of the presence of appreciable variations of the source polarizations.

## 4. CONCLUSIONS

We have derived integrated rotation measures for 18 sources, completing such RM measurements for all 34 BL Lac objects in the complete sample, using simultaneous VLA polarization observations at six frequencies below 2 GHz. The mean value of $|RM|$ for the entire sample is $\sim 40$ rad/m$^2$. The observed RM values are associated with thermal plasma external to the source, and probably arise when the radiation passes through the interstellar medium of the Galaxy.

However, such overall modest values for the integrated rotation measures do not rule out the presence of substantially higher RMs and sharp RM gradients in the inner regions of the compact VLBI jets of a few of these sources. Our multi-frequency VLBI polarization observations, currently being analyzed, will enable us to identify such cases of non-uniform Faraday rotation in the immediate vicinities of the compact radio sources.

## ACKNOWLEDGMENTS

The author thanks Denise Gabuzda for assistance with the primary calibration of the data, and also for valuable comments and constructive discussions of this work.